\documentclass{appolb}

\usepackage{fleqn,epsf,graphicx}
\usepackage{authblk}
\usepackage{cite}

\usepackage{epsfig}
\usepackage{graphicx}
\usepackage{caption}
\usepackage{subcaption}
\usepackage{hyperref}

\usepackage{amsmath}
\usepackage{xcolor}

\newcommand{\minus}{\scalebox{0.75}[1.0]{$-$}}

\begin{document}
\title{Introduction of total variation regularization into filtered backprojection algorithm
}
\author{

L.~Raczy\'nski$^{a}$, W.~Wi\'slicki$^{a}$, K.~Klimaszewski$^{a}$, W.~Krzemie\'n$^{b}$, P.~Kowalski$^{a}$,  R.~Shopa$^{a}$, 
P.~Bia\l{}as$^c$, C.~Curceanu$^d$, E.~Czerwi\'nski$^c$, K.~Dulski$^c$, A.~Gajos$^c$, B.~G\l{}owacz$^c$, M.~Gorgol$^e$, B.~Hiesmayr$^f$, B.~Jasi\'nska$^e$, 
D.~Kisielewska-Kami\'nska$^c$, G.~Korcyl$^c$, T.~Kozik$^c$, N.~Krawczyk$^c$, E.~Kubicz$^c$, M.~Mohammed$^{c,g}$, 
M.~Pawlik-Nied\'zwiecka$^c$, S.~Nied\'zwiecki$^c$, M.~Pa\l{}ka$^c$, Z.~Rudy$^c$, N.~G.~Sharma$^c$, S.~Sharma$^c$, M.~Silarski$^c$, M.~Skurzok$^c$, A.~Wieczorek$^c$, B.~Zgardzi\'nska$^e$, M.~Zieli\'nski$^c$, P.~Moskal$^c$

\address{
$^{a}$ Department of Complex Systems, National Centre for Nuclear Research, 05-400 Otwock-\'Swierk, Poland \\
$^{b}$ High Energy Physics Division, National Centre for Nuclear Research, 05-400 Otwock-\'Swierk, Poland \\
$^{c}$ Faculty of Physics, Astronomy and Applied Computer Science, Jagiellonian University, 30-348 Cracow, Poland \\
$^{d}$ INFN, Laboratori Nazionali di Frascati, 00044 Frascati, Italy \\ 
$^{e}$ Institute of Physics, Maria Curie-Sk\l{}odowska University, 20-031 Lublin, Poland \\
$^{f}$ Faculty of Physics, University of Vienna, 1090 Vienna, Austria \\
$^{g}$ Department of Physics, College of Education for Pure Sciences, University of Mosul, Mosul, Iraq
}

}

\maketitle
\begin{abstract}
In this paper we extend the state-of-the-art filtered backprojection (FBP) method with application of the concept of Total Variation regularization. We compare the performance of the new algorithm with the most common form of regularizing in the FBP image reconstruction via apodizing functions. The methods are validated in terms of cross-correlation coefficient between reconstructed and real image of radioactive tracer distribution using standard Derenzo-type phantom. We demonstrate that the proposed approach results in higher cross-correlation values with respect to the standard FBP method. 
\end{abstract}

  
\section{Introduction}

\begin{figure}[htb]
\centerline{%
\includegraphics[width=7cm]{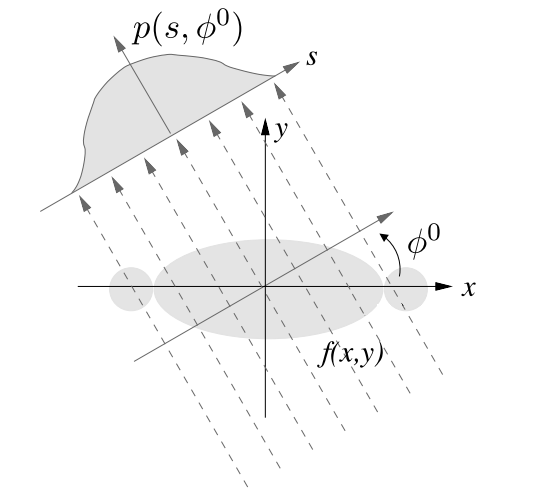}}
\caption{Data acquisition process in PET measurements. The figure is adapted from~\cite{Henkin}.}
\label{Fig:lors}
\end{figure}

Positron Emission Tomography (PET) is currently a key technique in the medical imaging area, which allows to diagnose functions of the organism and to track tumor changes. 
PET scanner consists of detector elements mounted on one or more rings, positioned so that it surrounds the patient~\cite{Karp, Slomka, Vander}. 
Those detectors are used to register pairs of gamma quanta emitted back-to-back from patient's body. 
In this work we will consider only a 2-dimensional PET scanner geometry. However, conclusions from this study may be easily extended to 3-dimensional case. The function $f(x, y)$ describes the radioactive tracer distribution (see Fig.~\ref{Fig:lors}). 
The measured data are collected as projections of the function $f(x,y)$ along the lines of response (LORs).
For instance, a projection $p(s, \phi^0)$ is formed by integration along all parallel LORs at a fixed angle $\phi^0$  
(see Fig.~\ref{Fig:lors}). 

The goal of the reconstruction process is to compute the unknown function $f(x,y)$ from registered collection of projections $p(s, \phi)$ for all angles $ 0 \leq \phi \leq \pi$.
However, the registered data $p(s, \phi)$ are stochastic while the inverse problem is ill-posed. For such case even a small perturbation of the data may lead to an unpredictable change in the resulting image $f$. Hence, additional constraints must be applied in order to ensure the computation of a meaningful solution. This is the essential goal of regularization methods. 

The filtered backprojection (FBP)~\cite{Hegl} was the first PET reconstruction technique and it is still treated as the reference method for more advanced approaches. In case of the FBP algorithm, the most common form of regularizing is via smoothing or so called apodizing functions~\cite{Henkin}.  

In this paper we propose a novel PET image reconstruction scheme based on the concept of total variation (TV) regularization~\cite{Rudin}. We investigate the quality
of image reconstruction based on the simulations of the Derenzo-type phantom with the J-PET detector~\cite{jpet_PM14, jpet_LR14, jpet_PM15, jpet_LR15, jpet_GK, jpet_JS, jpet_LR17}. 
We demonstrate that the TV regularization based image reconstruction algorithm performs better than the standard FBP algorithm with regularization via apodizing functions in terms of the quality of reconstructed images.

In Sec.~2 short description of the idea of regularization methods in PET image reconstruction is given. The state-of-the-art FBP algorithm is refined in order to apply the TV regularization method. The simulation of the J-PET tomograph as well as the results of comparative analysis of the two regularization techniques, TV method and apodizing functions, are presented in Sec.~3. The conclusions and directions for future work are presented in Sec.~4.

\section{Materials and methods}

\subsection{Filtered backprojection algorithm}

In the FBP algorithm the projection data are first filtered: 
\begin{equation}
	p^F (s, \phi) = \mathcal{F}^{-1}\left( W(v_s) |v_s| \mathcal{F}\left(  p(s, \phi) \right) \right)
	\label{Eq:FBP_step1}
\end{equation}
and then backprojected to image space 
\begin{equation}
	f(x, y) = \int_0^{\pi} p^F (s = x \cos \phi + y \sin \phi, \phi) d\phi.
	\label{Eq:FBP_step2}
\end{equation}
As shown in Eq.~(\ref{Eq:FBP_step1}), the filtering is performed in Fourier space with the ramp filter $|v_s|.$ 
Additionally, the characteristics of the high pass filter $|v_s|$ is multiplied with apodizing window $W(v_s).$ The purpose of the application of an apodizing window is to suppress amplification of high frequencies by the ramp filter $|v_s|,$ since the high frequency components of the projection data $\mathcal{F}(p(s, \phi))$ are dominated by noise. A very common form of apodizing function $W(v_s)$ is the Hamming or Hann window~\cite{Henkin}. During our preliminary studies, we observed that the best results are obtained for the Hann window.  


\subsection{Problem definition}

The application of the TV regularization to the FBP image reconstruction scheme requires an interchange of the order of the filtering and backprojection steps in 
Eqs.~(\ref{Eq:FBP_step1}) and (\ref{Eq:FBP_step2}). 
Hence, the projection data are first backprojected
\begin{equation}
	b(x, y) = \int_0^{\pi} p(s = x \cos \phi + y \sin \phi, \phi) d\phi
	\label{Eq:BPF_step1}
\end{equation}
and then filtered in image space. The filtering operation may be described by using the convolution equation
\begin{equation}
	b(x, y) = (h * f)(x,y).
	\label{Eq:BPF_step2}
\end{equation}
The original image $f(x,y)$ is convolved with impulse response of the ramp filter $h(x,y)$ to produce observed, backprojected image $b(x,y).$
The filtering can be performed via the deconvolution of $b(x,y)$ with impulse response $h(x,y).$ The impulse response $h$ may be rewritten as a cyclic and square matrix $A$ and hence the filtration formula may be rewritten to the matrix notation
\begin{equation}
	\mathbf{b} = A \mathbf{f}.
	\label{Eq:bAf}
\end{equation}
The bold symbols $\mathbf{b}$ and $\mathbf{f}$ in Eq.~(\ref{Eq:bAf}) represent the vectorized versions of the functions $b(x,y)$ and $f(x,y),$ respectively. 
The inverse problem defined in Eq.~(\ref{Eq:bAf}) is ill-posed and the regularization methods are required in order to calculate a meaningful solution, as mentioned in the introduction.

\subsection{Total Variation regularization}

The most common class of regularization methods in image processing is based on TV approach~\cite{Blom, Brook}. The TV of image 
$f$ is the sum of the magnitudes of its discrete gradient at every pixel
\begin{equation}
\operatorname{TV}(f) = \sum_x \sum_y |Df(x,y)|
\end{equation}
where $D$ is a gradient operator. The reconstruction algorithm finds the solution of Eq.~(\ref{Eq:bAf}) by solving the unconstrained problem:
\begin{equation}
 	\min_f ~\left(\operatorname{TV}(f) + \frac{\mu}{2}\|A\mathbf{f}- \mathbf{b}\|_2^2  \right),
  	\label{Eq:TV_uncons} 
\end{equation}  
where $\mu$ is the regularization parameter. This approach would succeed if the gradient of the underlying image is sparse~\cite{Candes, Donoho}. Hence, TV algorithm can reconstruct not only sparse images but also dense piecewise constant images. The theory for penalty functions implies that the solution in Eq.~(\ref{Eq:TV_uncons}) approaches the solution of Eq.~(\ref{Eq:bAf}) as $\mu$ approaches infinity. To solve the problem in Eq.~(\ref{Eq:TV_uncons}), we apply the augmented Lagrangian algorithm~\cite{Hest, Chan}.

We wish to make a comment about the related works. PET reconstruction using TV regularization was investigated by several groups~\cite{Barbano, Buletin, Inv}. However, typically the reconstruction problem takes into account only the information about the projection data $p(s, \phi)$ and not the backprojected image $b(x,y)$ as in Eq.~(\ref{Eq:TV_uncons}). Hence, in the proposed scenario all the optimization process, defined in Eq.~(\ref{Eq:TV_uncons}), is performed in the image space and not in the projection space. One of the benefits of processing in the image space is the possibility to include the information from Time of Flight (TOF) measurement~\cite{Karp, Kardmas}. Hence, a presented approach may be extended to PET scanners that provide TOF measurement.   

\section{Results}

For the evaluation of the two regularization approaches, the TV method and apodizing functions technique, projection data of a Derenzo-type phantom have been used. The Derenzo-type phantom consists of sets of rods with diameters 10, 15, 23, 32, 40 and 48~mm and the same separation between surfaces in the corresponding sets. Since only a 2-dimensional geometry of the PET tomograph has been studied, the length of the rods was set to zero. 

Sample data were produced using Monte Carlo simulation of the J-PET scanner with 384 strips arranged in one 
layer~\cite{jpet_PK15, jpet_PK16}. The J-PET detector was defined as a cylinder with inner radius 428~mm. The phantom was placed in the plane $(x,y),$ which is perpendicular to the J-PET tomograph main axis, in its center position $(z = 0).$ During the reconstruction process we considered only the projections from the $z = 0$ plane and we collected $10^6$ coincidence events in total. The reconstructed images were 512 pixels $\times$ 512 pixels and the pixel size was  1~mm $\times$ 1~mm. 

As the quality measurement, we selected the cross-correlation coefficient calculated between the real and reconstructed image.  The cross-correlation coefficient $\rho$ was calculated according to the equation
\begin{equation}
 	\rho = \frac{\sum_{x=1}^{512} \sum_{y=1}^{512} ( \hat{f}(x,y) - \hat{f}_m)( f(x,y) -f_m)}{\sqrt{\sum_{x=1}^{512} \sum_{y=1}^{512} ( \hat{f}(x,y) - \hat{f}_m)^2
\sum_{x=1}^{512} \sum_{y=1}^{512} (f(x,y) - f_m)^2}}
  	\label{Eq:rho} 
\end{equation}  
where $\hat{f}(x,y)$ and $f(x,y)$ are the reconstructed and true images, respectively, and $\hat{f}_m$ and $f_m$ are the reconstructed and true images mean values, respectively. The values of $\rho$ are in the range from $\minus 1$ to 1, where value 1 corresponds to fully correlated images.

\begin{figure}[htb]
\centerline{%
\includegraphics[width=10cm]{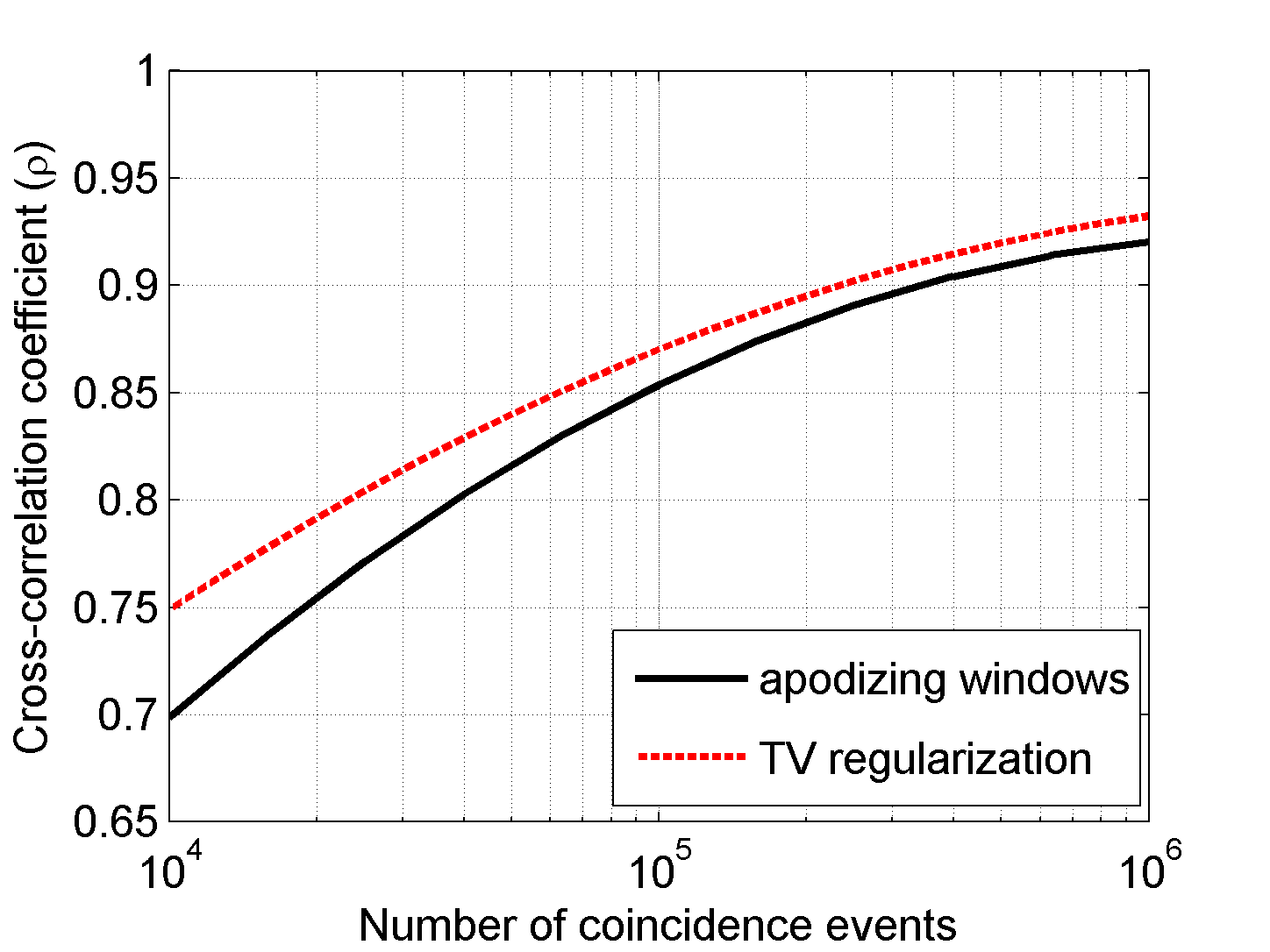}}
\caption{Performance of image reconstruction methods as a function of number of coincidence events.}
\label{Fig:corrcoef_results}
\end{figure}

Both reconstruction methods were optimized in the sense of choosing the regularization parameters that maximize the calculated value of cross-correlation coefficients. The results of the optimization process for both regularization methods are presented in Fig.~\ref{Fig:corrcoef_results}. 

The calculated cross-correlation coefficients for the image reconstruction method based on TV regularization are marked with dashed red line in Fig.~\ref{Fig:corrcoef_results} and take higher values in a wide range of number of coincidence events from $10^4$ to $10^6$ than reconstruction via apodizing windows (black line in Fig.~\ref{Fig:corrcoef_results}). 
The difference between those two functions is larger for small numbers of coincidence events and is about $8\%$ for $10^4$ events. From Fig.~\ref{Fig:corrcoef_results} it may be seen that both reconstruction methods have the same asymptotic properties; as the number of coincidence events goes to infinity, the cross-correlation coefficient goes to 1. 
However, in the case of TV regularization-based approach, the convergence process is faster. 

\begin{figure}[htb]
\centerline{%
\includegraphics[width=13cm]{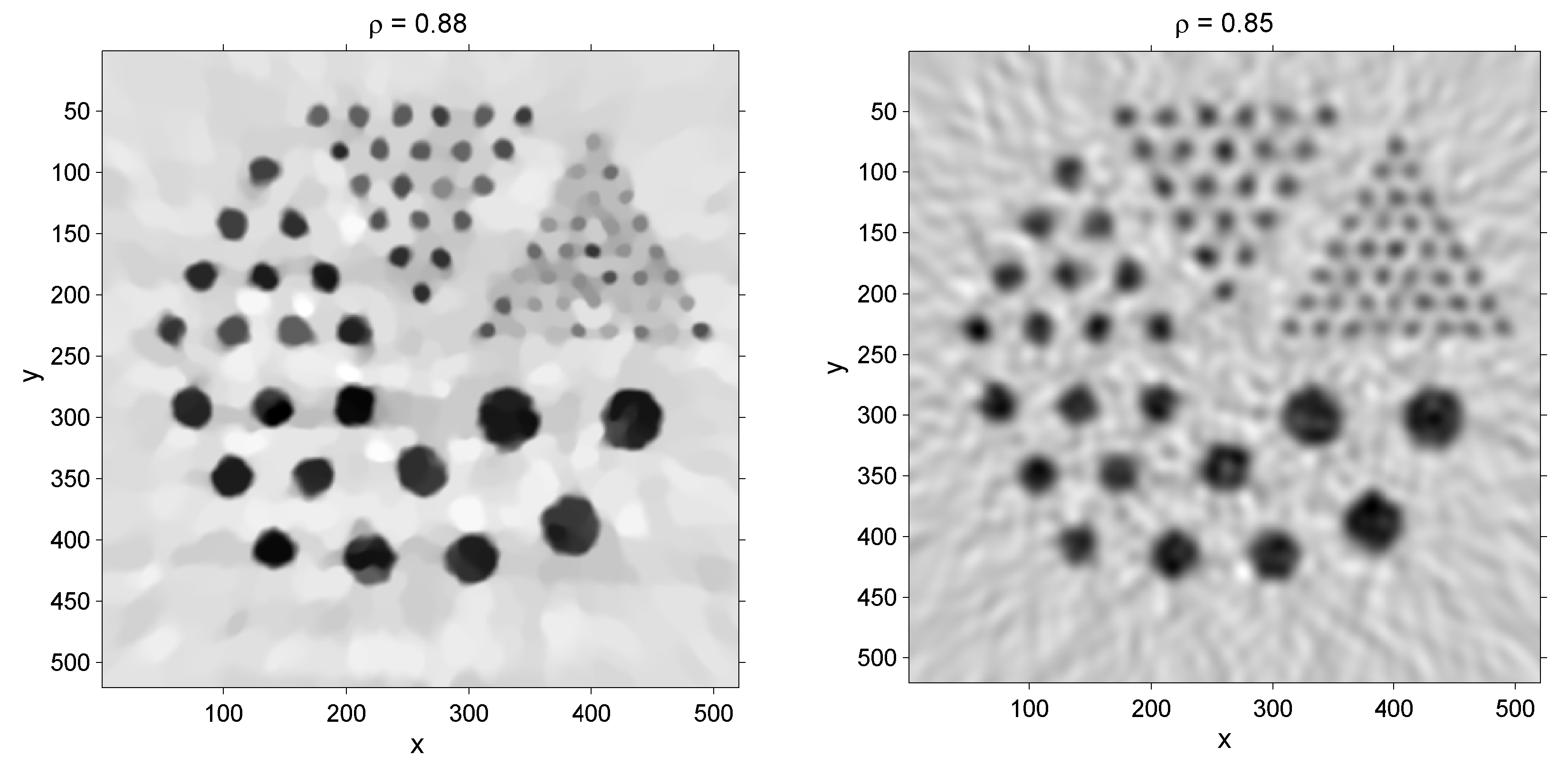}}
\caption{Images reconstructed with TV regularization algorithm (left)  and apodizing window function (right). In both cases the number of selected events was the same and equal to $10^5.$}
\label{Fig:examples}
\end{figure}


The two image reconstruction examples, based on TV regularization method and apodizing functions, are shown in Fig.~\ref{Fig:examples} on the left and right side, respectively.   
In both cases the number of selected events was the same and equal to $10^5.$ The cross-correlation coefficient evaluated between the real and reconstructed image in case of regularization based on TV method and apodizing functions was equal to 0.88 and 0.85, respectively. 

\section{Conclusions}

In this paper a novel scheme of regularization in PET image reconstruction, based on the TV method was introduced. 
We have shown that the use of the TV regularization method instead of the most common regularizing approach via apodizing windows improves the quality of reconstructed images. The calculated cross-correlation coefficients for the image reconstruction method based on TV regularization take higher values in a wide range of number of coincidence events from $10^4$ to $10^6$ than reconstruction via apodizing windows.

Future work will address other aspects of the proposed image processing scheme. One of the benefits of the processing in image space according to Eq.~(\ref{Eq:TV_uncons}) is that 
the information from TOF measurement may be easily included in the image reconstruction. The PET image reconstruction algorithm which additionally allows to take into account the TOF information is extremely important as the novel PET scanners tend to improve time resolution~\cite{jpet_PM16, Lecoq, Lecoq2, Varun}.  

\section*{Acknowledgement}

We acknowledge support by the National Science Centre through the grant No. 2016/21/B/ST2/01222.

\end{document}